\documentclass[cits]{PoS}

\usepackage[mathscr]{euscript}
\usepackage{amsmath,amsfonts,amssymb}

\usepackage{dcolumn}
\usepackage{graphicx}

\def\chpt{\raise0.4ex\hbox{$\chi$}PT}
\def\schpt{S\raise0.4ex\hbox{$\chi$}PT}
\def\rschpt{rS\raise0.4ex\hbox{$\chi$}PT}
\def\ie{{\it i.e.},\ }

\def\cO{{\cal O}}

\def\eq#1{Eq.~(\ref{eq:#1})}

\def\aschpt{HMrAS\raise0.4ex\hbox{$\chi$}PT}

\newcommand{\pole}{{\text{pole}}}

\newcommand{\lat}{{\text{lat}}}

\newcommand{\RS}{{\text{RS}}}

\newcommand{\MSbar}{{\overline{\text{MS}}} }
\newcommand{\Lambdabar}{{\overline{\Lambda}} }
\newcommand{\cst}{{c^*}}

\newcommand{\mcstbar}{\overline{m}_{c^*}}

\newcommand{\be}{\begin{equation}}
\newcommand{\ee}{\end{equation}}
\newcommand{\bea}{\begin{eqnarray}}
\newcommand{\eea}{\end{eqnarray}}
\newcommand{\mvt}{m_\text{v}}
\newcommand{\vt}{\text{v}}
\newcommand{\MHsinv}{\frac{\Lambda_{\rm  HQET}}{M_{H_s}}}

\newcommand{\Minv}{\frac{\Lambda_{\rm  HQET}}{M}}

\title{\vspace{-1mm}
Decay constants $f_B$ and $f_{B_s}$ and quark masses $m_b$ and $m_c$ from HISQ simulations 
\vspace{-0.5mm} }

\ShortTitle{\vspace{-2mm}Decay constants $f_B$ and $f_{B_s}$ and quark masses $m_b$ and $m_c$}

\author{\vspace{-4mm}
\speaker{J.~Komijani}$^{\;a,b}$,
A.~Bazavov$^c$,
C.~Bernard$^d$,
N.~Brambilla$^{a,b}$\thanks{Sec.~{3} only.},
N.~Brown$^d$,
C.~DeTar$^e$,
D.~Du$^f$,
A.X.~El-Khadra$^{g,h}$,
E.D.~Freeland$^i$,
E.~G\'amiz$^j$,
Steven~Gottlieb$^k$,
U.M.~Heller$^l$,
A.S.~Kronfeld$^{b,h}$,
J.~Laiho$^f$, 
P.B.~Mackenzie$^h$,
C.~Monahan$^e$\thanks{Present address: Department of Physics, Rutgers University, New Brunswick, NJ 089001, USA.},
Heechang~Na$^e$\thanks{Present address: Ohio Supercomputer Center, Columbus, OH 43215, USA.},
E.T.~Neil$^m$,
J.N.~Simone$^h$,
R.L.~Sugar$^n$,
D.~Toussaint$^o$,
A.~Vairo$^a\overset{\dagger}{,}$
R.S.~Van~de~Water$^h$
\\
\llap{$^a$} Physik-Department, Technische Universit\"at M\"unchen, Garching 85748, Germany\\
\llap{$^b$} Institute for Advanced Study, Technische Universit\"at M\"unchen, Garching 85748, Germany\\
\llap{$^c$} Department of Computational Mathematics, Science and Engineering
and Department of Physics and Astronomy, Michigan State University, East Lansing, MI 48824, USA \\
\llap{$^d$} Department of Physics, Washington University, St. Louis, MO 63130, USA\\
\llap{$^e$} Department of Physics and Astronomy, University of Utah, Salt Lake City, UT 84112, USA\\
\llap{$^f$} Department of Physics, Syracuse University, Syracuse, NY 13244, USA\\
\llap{$^g$} Department of Physics, University of Illinois, Urbana,  IL 61801, USA\\
\llap{$^h$} Fermi National Accelerator Laboratory\thanks{Operated by Fermi Research Alliance, LLC, 
under Contract No.~DE-AC02-07CH11359 with the US DOE.}, Batavia, IL 60510, USA\\
\llap{$^i$} Liberal Arts Department, School of the Art Institute of Chicago, Chicago, IL 60603, USA\\
\llap{$^j$} CAFPE and Departamento de F\'isica Te\'orica y del Cosmos, Universidad de Granada, E-18071 Granada, Spain\\
\llap{$^k$} Department of Physics, Indiana University, Bloomington, IN 47405, USA\\
\llap{$^l$} American Physical Society, One Research Road, Ridge, NY 11961, USA\\
\llap{$^m$} Department of Physics, University of Colorado, Boulder, CO 80309, USA\\
\llap{$^n$} Department of Physics, University of California, Santa Barbara, CA 93106, USA\\
\llap{$^o$} Physics Department, University of Arizona, Tucson, AZ 85721, USA

\vspace{1mm}
\textbf{\textsf{Fermilab Lattice, MILC and TUMQCD Collaborations}}
\vspace{1mm}

E-mail:
\email{j.komijani@tum.de}, \email{ask@fnal.gov}, \email{cb@wustl.edu}, \email{doug@physics.arizona.edu}
}
\vspace{-2mm}

\abstract{
We present a progress report on our calculation of the decay constants
$f_B$ and $f_{B_s}$ from lattice-QCD simulations with highly-improved staggered quarks.
Simulations are carried out with several heavy valence-quark masses on 
$(2+1+1)$-flavor ensembles that include charm sea quarks.
We include data at six lattice spacings and several light sea-quark masses, 
including an approximately physical-mass ensemble at all but the smallest lattice spacing, 0.03 fm. 
This range of parameters provides excellent control of the continuum extrapolation to zero lattice spacing
and of heavy-quark discretization errors. 
Finally, using the heavy-quark effective theory expansion 
we present a method of extracting from the same correlation functions
the charm- and bottom-quark masses as well as some low-energy constants
appearing in the heavy-quark expansion. 
}

\FullConference{34th annual International Symposium on Lattice Field Theory\\
		24-30 July 2016\\
		University of Southampton, UK}

\begin{document}

\section{Introduction}
\vspace{-2mm}
One way of searching for new physics is by looking for discrepancies between 
precise experimental measurements and equally precise theoretical calculations within the Standard Model (SM).
To this end, the study of heavy-light mesons provides a rich area for investigation. 
B-meson decay constants enter SM predictions for leptonic decays of charged and neutral B mesons. 
The former can be used to determine the corresponding CKM elements, while the latter, being loop-induced in the SM 
provides a window for New Physics. 
The study of heavy-light meson masses within the framework of heavy quark effective theory 
(HQET) enables determinations of charm- and bottom-quark masses and also some low-energy constants (LECs) appearing in 
HQET, which in turn can be used for inclusive determinations of $|V_{ub}|$ and $|V_{cb}|$. 

Here we provide a progress report on our effort to calculate the leptonic decay 
constants $f_B$ and $f_{B_s}$ in four-flavor lattice QCD~\cite{Lattice:2015nee}.
The calculation employs highly improved staggered quarks 
(HISQ)~\cite{HPQCD_HISQ,milc_hisq,scaling09,HISQ_CONFIGS} 
with masses heavier than the charm-quark mass. 
For details about the method for extracting the decay constants 
from two-point correlation functions, see Ref.~\cite{Bazavov:2014wgs}. 
We also extend this work to study the masses of heavy-light mesons. 
Within the framework of HQET, we present a method to organize the heavy-quark mass dependence of heavy-light mesons. 
This method leads to lattice-QCD determinations of quantities such as $\Lambdabar$ and $\mu_\pi^2$ that appear in HQET. 
Equivalently, this method provides a way to calculate the charm- and bottom-quark masses. 

\begin{figure}[bp]
\begin{tabular}{c c}
\includegraphics[width=0.4\textwidth]{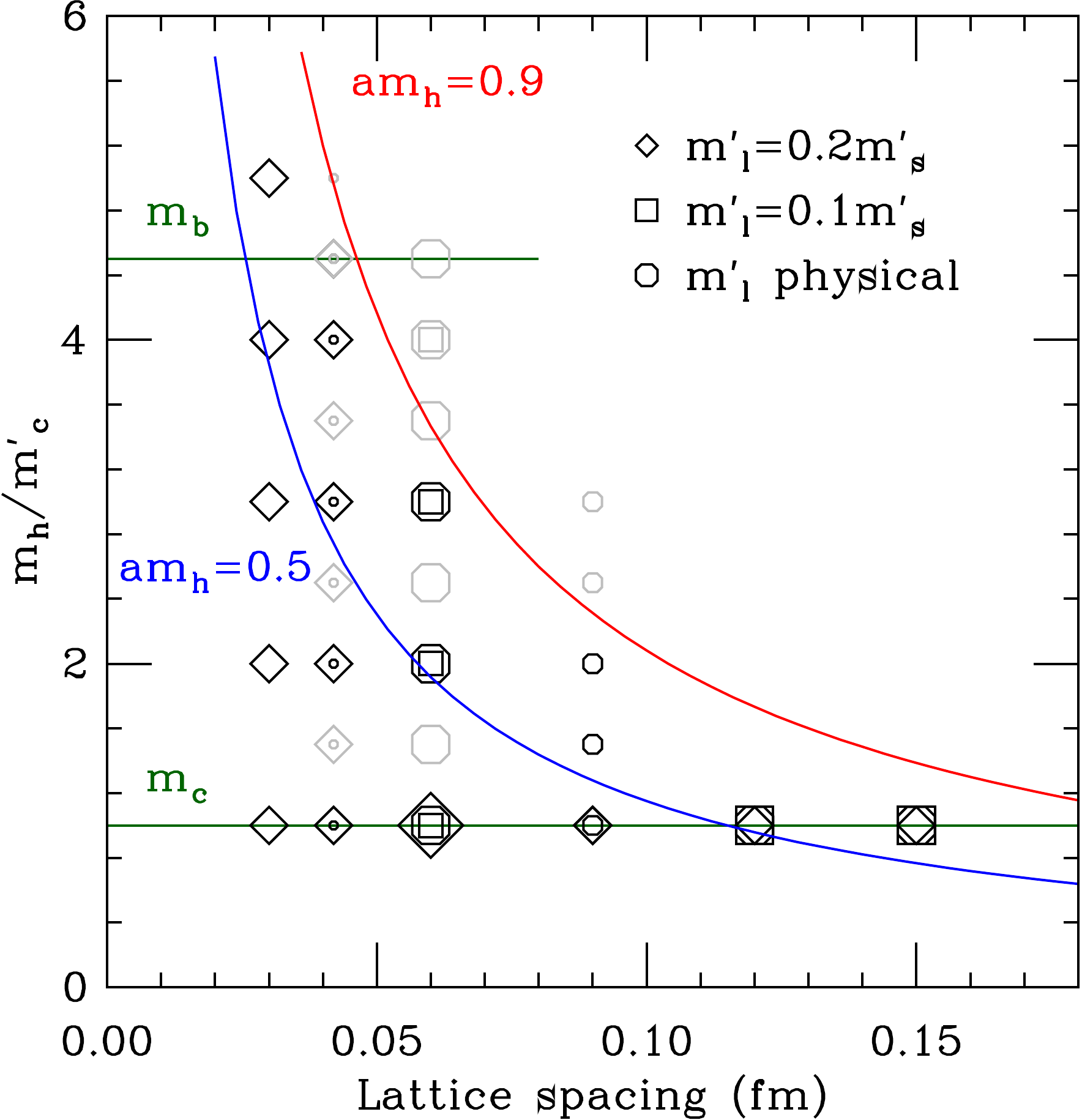} &
\hspace{1cm} \includegraphics[width=0.42\textwidth, trim={0 0.5cm 0 0},clip]{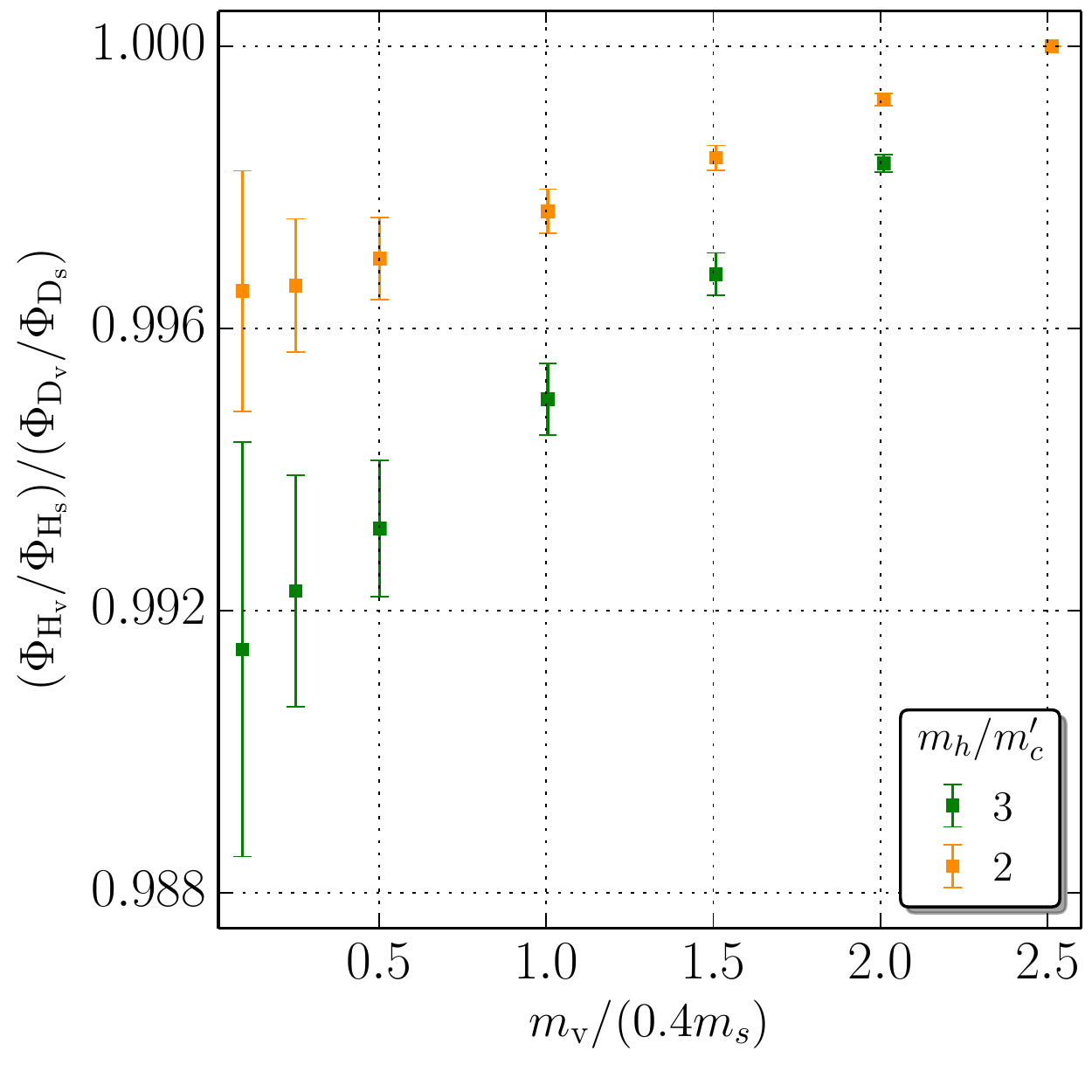}
\end{tabular}
\caption{\label{fig:1} 
Left: valence heavy-quark masses and lattice-spacings of ensembles in this study for 
different light-to-strange sea-quark-mass ratios $m'_l/m'_s$. 
(Primes on the masses indicate the simulation mass values.)
The symbol radius is proportional to the data sample size. 
The red line indicates the cut $am_h = 0.9$; the blue, $am_h = 0.5$. 
Right: double ratio of decay constants for the physical-mass ensemble 
at 0.06~fm as a function of valence light-quark mass $\mvt$. 
In this double ratio the leading-order terms in HQET and \schpt\ cancel, 
revealing the higher-order terms 
that depend upon both light- and heavy-quark masses.
}
\end{figure}

The range of valence heavy-quark masses and lattice-spacings for the QCD gauge-field ensembles 
in this study is shown in the left panel of Fig.~\ref{fig:1}. 
Compared with Ref.~\cite{Lattice:2015nee}, our analysis now includes an $a\approx0.03$~fm, $m'_l/m'_s = 0.2$, 
ensemble for which $am_b \approx 0.6$, and thus no extrapolation from lighter heavy-quark masses is needed. 
In order to avoid large lattice artifacts we drop data with $am_h>0.9$ 
and use a parameterization  of the heavy-quark mass dependence guided by HQET in our chiral-continuum fits.

\vspace{-2mm}
\section{Chiral-continuum-HQET fit of decay constants}
\label{sec:decay-constants_chiral-analysis}
\vspace{-2mm}

We use HQET to model the heavy-quark mass dependence of the decay constants. 
In heavy-quark physics,
the conventional pseudoscalar-meson decay constant is $\Phi = f \sqrt{M}$. 
Let us start with massless light quarks. The decay constant in this limit, denoted by $\Phi_0$, can be expanded as
\bea
  \Phi_0 &=& C \, \left(1 + k_1 \Minv + k_2 \bigl(\Minv\bigr)^2 + \cdots \right) \tilde{\Phi}_0\, ,
    \label{eq:decay-constant:chiral-limit}
\eea
where $\tilde{\Phi}_0$ is the matrix element of the HQET current in the 
infinite-mass limit
and the Wilson coefficient $C$ arises from matching the QCD current and the HQET current
\begin{equation}
   C = \Bigl[ \alpha_s(m_Q)\Bigr]^{-2/\beta_0} \Bigl(1 + {\cal O}(\alpha_s) \Bigr) \, ,
   \label{eq:Wilson-coefficient}
\end{equation}
where $m_Q$ is the heavy-quark mass and $\beta_0 = 11 - 2n_f/3 = 25/3$ in our simulations.\footnote{Here,  
$\tilde{\Phi}_0$ is a renormalization-group invariant quantity, 
\ie the renormalization scale and scheme dependence of the HQET current in the infinite-mass limit 
and the Wilson coefficient $C$ cancel. 
Consequently, $C$ in \eq{Wilson-coefficient} does not depend on the scale of the effective current.} 
Within the framework of {heavy-meson, rooted, all-staggered chiral perturbation theory}
(HMrA\schpt)~\cite{Bernard-Komijani},  
this relation can be extended to include the light-quark mass dependence and taste-breaking discretization errors. 
This provides a suitable fit function to perform a combined chiral-continuum fit to lattice data at multiple lattice 
spacings and various valence- and sea-quark masses. 

The fit function that we use in this analysis has the following schematic form
\begin{equation}
\Phi_{H_\vt} = 
  C\, \left(1 + k_1 \MHsinv + k_2 (\MHsinv)^2 + k'_1\frac{m_c}{m'_c}\right)\,
    \Bigl(1+ \text{log/analytic terms}\Bigr)
    \left(\frac{m'_c}{m_c}\right)^{3/27} \tilde{\Phi}_0 \, ,
    \label{eq:schematic:function}
\end{equation}
where the ``log/analytic" terms include the next-to-leading order (NLO) staggered chiral logarithms, 
and the NLO, NNLO, and NNNLO analytic terms in the valence and sea-quark masses. 
The NLO staggered chiral logarithms are given in equation 177 of Ref.~\cite{Bernard-Komijani},  
and the NLO analytic terms (at a fixed heavy-quark mass) are 
\bea
 L_\text{s}\, (2m_l+m_s) + L_\text{v}\,\mvt + L_{a}\,a^2\, .
\eea
Because we have a wide range of heavy-quark masses from near charm to bottom (at the finest lattice spacings)
it is important to take the heavy-quark mass dependence of $L_\text{s}$ and $L_\text{v}$ into account. 
The importance of this dependence is shown in the right panel of Fig.~\ref{fig:1}, 
where a double ratio of decay constants $(\Phi_H/\Phi_{H_s})/(\Phi_D/\Phi_{D_s})$
is constructed to be sensitive to higher-order terms that depend upon both the light- and heavy-quark masses.
Based on the observed quark-mass dependence of the double ratio, we allow the NLO analytic-term LECs 
to have $M_{H_s}$ dependence, replacing
\bea 
 L_\text{s} &\to& L_\text{s} + L'_\text{s} \MHsinv + L''_\text{s} (\MHsinv)^2 \, ,\\
 L_\text{v} &\to& L_\text{v} + L'_\text{v} \MHsinv + L''_\text{v} (\MHsinv)^2 \, .
\eea
Because we have very precise data, the NLO terms in HMrA\schpt\ are insufficient to describe the quark-mass dependence. 
We therefore include in our preferred fit all NNLO and NNNLO purely mass-dependent analytic terms, but omit terms that 
mix powers of lattice spacing and light-quark masses. 
Similar to $L_\vt$ and $L_\text{s}$, the LECs appearing at higher orders in principle have heavy-quark mass 
corrections, although in practice most of these corrections are not needed to obtain an acceptable fit. 
A heavy-quark mass dependence also appears implicitly in the NLO chiral logarithms
through the $M_{H_\vt^*} - M_{H_\vt}$ hyperfine splitting and heavy-light flavor splittings.
The factor $(m'_c/m_c)^{3/27}$ in \eq{schematic:function} incorporates the leading effect of mistunings 
in the simulated sea charm-quark mass $m'_c$ compared to the physical charm mass $m_c$. 
The coefficient $\tilde{\Phi}_0$ in \eq{schematic:function}, which is a constant in the continuum limit, 
has a generic lattice-spacing dependence, which we parameterize as
\vspace{-0.1cm}
\be
 \tilde{\Phi}_0\,\to\,
    \left( 1 + c_1 \alpha_s\, (a\Lambda)^2 + c_2(a\Lambda)^4 + c_3 \alpha_s\, (am_h)^2 
    + c_4(am_h)^4 + {c_5 \alpha_s\, (am_h)^4} \right)\, \tilde{\Phi}_0 \, .
\ee 
The LECs appearing at NLO and higher orders also have lattice artifacts, 
but we do not incorporate them into our fit here.  

\begin{figure}[bp] 
\begin{tabular}{c c}
\includegraphics[width=0.53\textwidth, trim={0 0 0 0.1cm},clip]{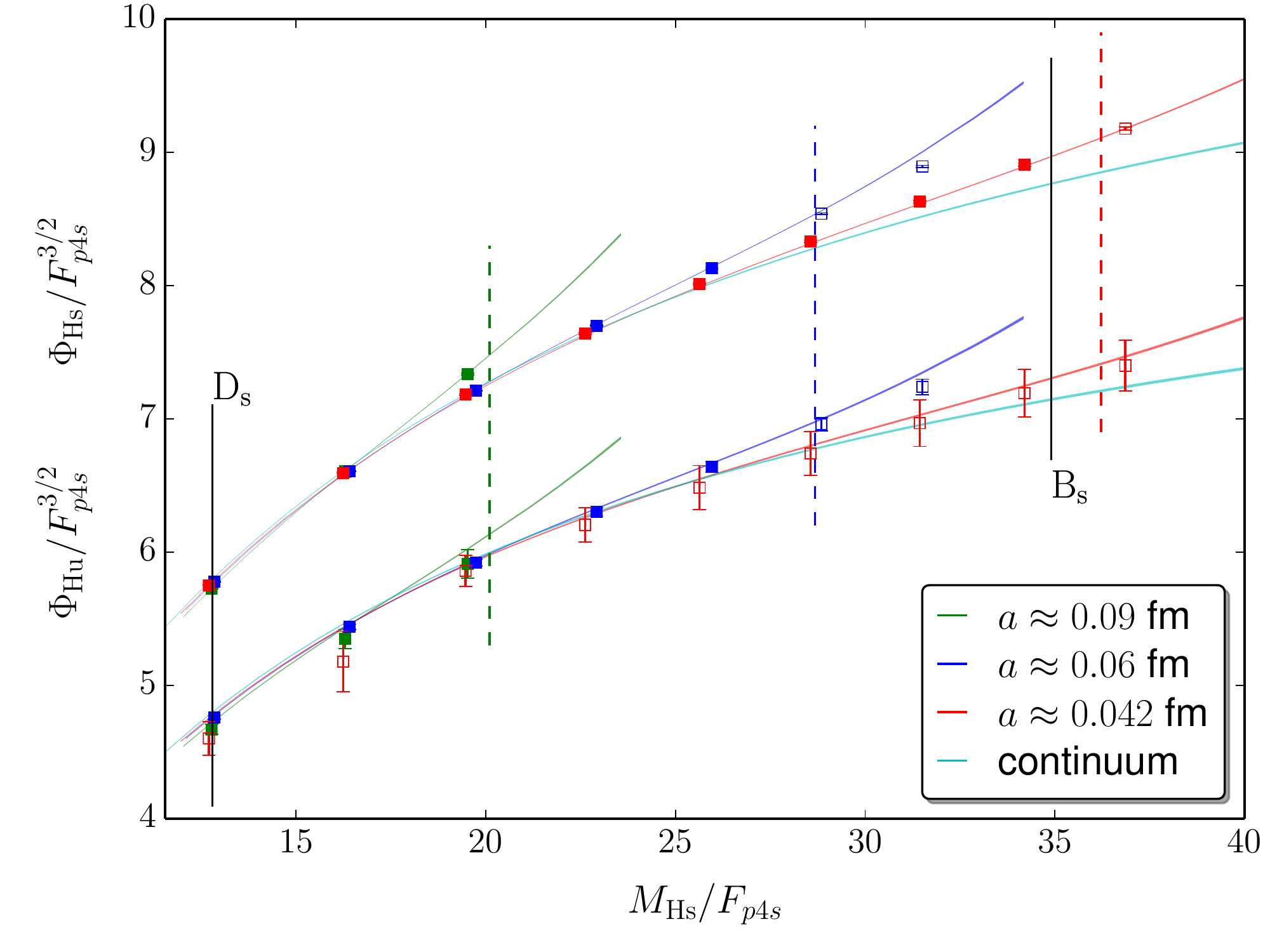} &
\includegraphics[width=0.42\textwidth, trim={14cm 0 0 0.1cm},clip]{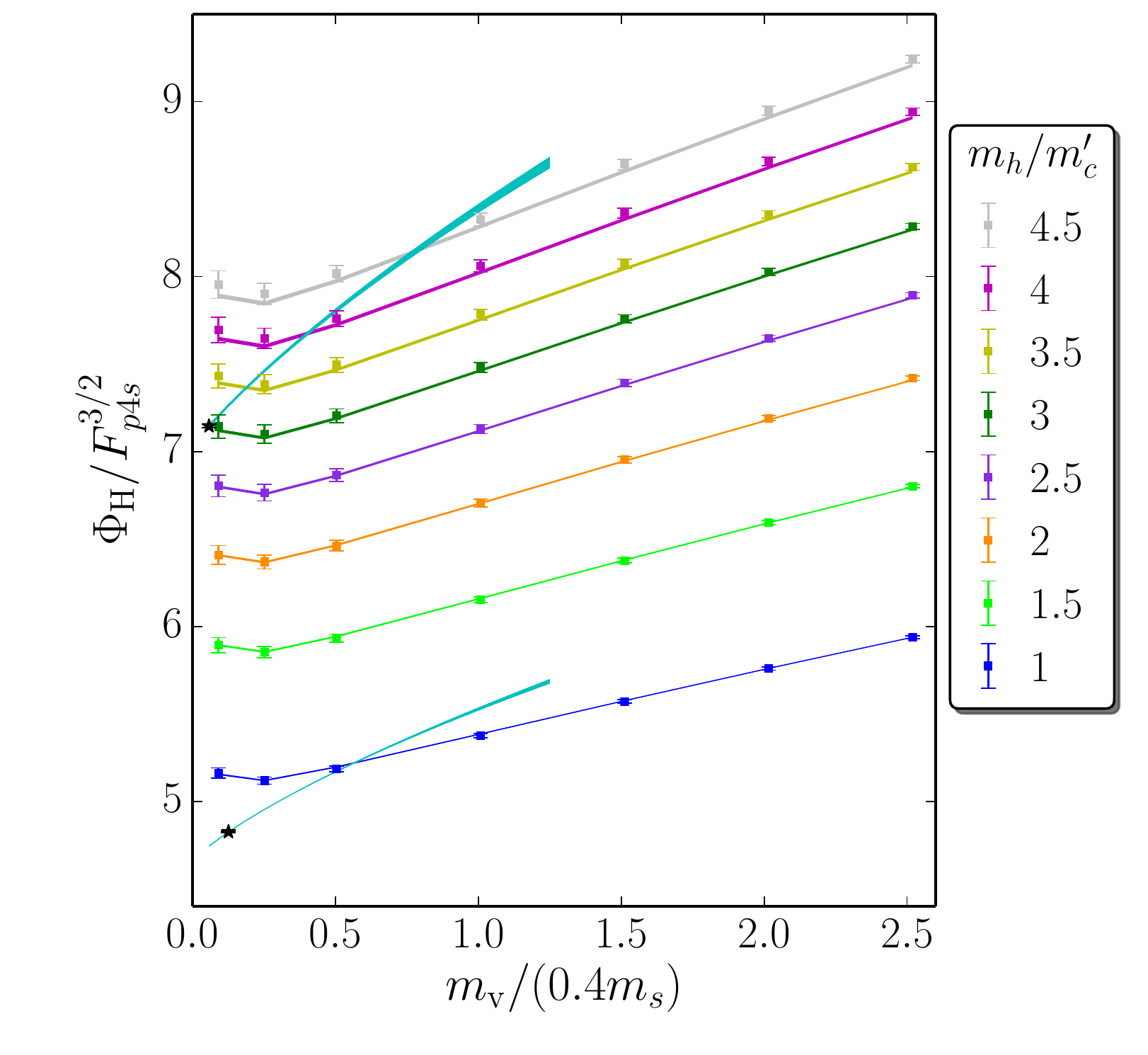} 
\end{tabular}
\vspace{-5mm}
\caption{\label{fig:Decays}
Chiral-continuum-HQET fit of decay-constant data at 6 lattice spacings with valence heavy-quark masses  
in the range $am'_c\le am_h<0.9$. This fit has a correlated $\chi^2/\text{dof} = 265/305$, giving $p = 0.95$. 
Left: decay constants plotted in units of $F_{p4s}$ vs. the heavy-strange meson mass for three lattice 
spacings and the continuum extrapolation 
(see Ref.~\cite{Bazavov:2014wgs} for the definition and determination of the scale $F_{p4s}$ on the HISQ ensembles). 
Data points to the right of the dashed vertical line of the corresponding color are excluded from the fit. 
The open symbols indicate the data points that are omitted. 
Right: decay constants for the $a \approx 0.042$~fm, $m'_l/m'_s = 0.2$ data vs. the valence light-quark mass.
The full-QCD, continuum-limit results at physical $c$- and $b$-quark masses are shown in cyan. 
The stars indicate the physical light-quark-mass results for the $B^+$ and $D^+$ mesons.
}
\end{figure}

In total, our base fit function in this analysis has 23 parameters. 
Figure~\ref{fig:Decays} shows two projections of the resulting fit. 
From the fit function evaluated at zero lattice spacing and physical sea-quark masses, 
we obtain the decay constants as a function of $M_{H_s}$ and the valence light-quark mass. 

\vspace{-2mm}
\subsection{Error budget}
\label{sec:conclusions}
\vspace{-2mm}
To estimate the systematic errors on the decay constants,   
we rerun the analysis with alternative fit functions, including or dropping the coarsest ensembles, 
and with various choices for scale-setting quantities and tuned quark masses. 
(For details see Refs.~\cite{Lattice:2015nee,Bazavov:2014wgs}.) 
After rejecting the fits with $p<0.05$, we take the extremes of the histograms  
as our estimate of the systematic error from the chiral-continuum-HQET fit. 
Preliminary error budgets for $f_B$ and $f_{B_s}$ are presented in Table~\ref{tab:Error-Budget}. 
Compared with our previous report~\cite{Lattice:2015nee}, 
the uncertainty has been decreased, primarily because our analysis now includes more data at 0.042~fm
and an even finer ensemble at 0.03~fm. 

\begin{table}[tp]
\caption{
  Preliminary error budgets. 
  The first error is the statistical error from our base fit. 
  The second error, determined from histograms, is our estimate of the fit systematic error. 
  The third error comes from propagating the error of the PDG value of $f_\pi$, 
  which is the physical quantity used to set the lattice scale. 
}\label{tab:Error-Budget}
\centering
\begin{tabular}{lcc}
    \hline \hline
	& $f_{B^+}$ MeV & $f_{B_s}$ MeV\\
    \hline
    Statistics  & 0.4 & 0.3 \\
    Alternative fit choices & 1.1 & 0.8 \\
    $f_\pi$  & 0.3 & 0.4 \\
    \hline
    Total  & 1.2 & 1.0 \\
    \hline \hline
  \end{tabular}
\end{table}

\vspace{-2mm}
\section{HQET and heavy-light meson masses}
\label{sec:MesonMasses}
\vspace{-2mm}
Heavy-light meson masses are given in the HQET as an expansion in the heavy-quark mass~$m_Q$: 
\begin{eqnarray}
    M_H  &=&  m_Q + \bar{\Lambda} + \frac{\mu_\pi^2 - \mu_G^2(m_Q)}{2m_Q}  
    + \cO\left( \frac{1}{m_Q^{2}} \right)\,. \label{eq:PS_Mass:HQET:mQ} 
\end{eqnarray}
The LECs appearing in this relation have a simple physical interpretation: $\bar{\Lambda}$ is the energy of
the light quarks and gluons; $\mu_\pi^2/2m_Q$ is the kinetic energy of the heavy quark; and $\mu_G^2/2m_Q$
corresponds to the hyperfine energy of the interaction between the heavy quark's spin and the chromomagnetic
field inside the meson, $\mu_G$ depends on the heavy-quark mass through the chromomagnetic Wilson
coefficient, which is known to three loops~\cite{Grozin:2007fh}.
These LECs also appear in the heavy-quark expansions of inclusive semileptonic decay rates.
Because of the nonperturbative nature of the LECs, lattice gauge theory is crucial for their
determination~\cite{Kronfeld:2000gk}.

In order to use \eq{PS_Mass:HQET:mQ} as a fitting function for the different values of heavy valence-quark masses 
for which we have data, we have to give a precise meaning to the heavy-quark mass $m_Q$. 
The first step is to relate the bare lattice quark mass to a continuum mass.
The one-loop relation between the bare lattice mass of a heavy quark $h$, $am_h$, 
with its pole mass, $m^\pole_h$, for staggered quarks is~\cite{Mason:2005bj}
\bea
  m^\pole_h &=& \frac{am_h}{a}
  \left\{ 1 + \alpha_\lat\left[-\frac{2}{\pi}\log(a m_h)+A_{10}\right] + \cO(\alpha_\lat^2)\right\}\, , 
  \label{eq:bare2pole:lattice} 
\eea
where  $A_{10} = K_0 + K_1 (am_h)^2 + \cdots$, and $K_1$ is an unknown constant.
The relation between the pole mass and the $\MSbar$ mass is known up to order $\alpha_s^4$~\cite{Marquard:2016dcn}.  
Expressing the pole-mass in terms of the $\MSbar$-mass and using \eq{bare2pole:lattice} we can relate the $\MSbar$ mass to the bare lattice mass:
\bea
  \frac{m_h^\MSbar(\mcstbar)}{\mcstbar}
   &=& \frac{am_h}{am_\cst} 
  \Big\{ 1 + \alpha_\MSbar(\mcstbar)\, 
  \left[K_1\bigl((a m_h)^2 - (a m_\cst)^2\bigr) + \cdots \right]
  + \cdots \Big\} \,,
\label{eq:MSbarratio2bareratio:lattice:hc} 
\eea
where $a m_\cst$ denotes a reference mass, which in our analysis will be a tuned charm-quark mass.
The dots stand for higher-order terms in the coupling and in the lattice spacing.
We have set the scale of the $\MSbar$ mass to be $\mcstbar = m_\cst^\MSbar(\mcstbar)$. 
We are currently treating $\mcstbar$ as a free parameter in this analysis, 
allowing the fit to absorb perturbative and scale errors coming from the relation between $\mcstbar$ 
and $a m_{c^*}$.
The lattice artifacts appearing on the right-hand side of \eq{MSbarratio2bareratio:lattice:hc} vanish in the continuum 
limit.

We could interpret the heavy-quark mass appearing in \eq{PS_Mass:HQET:mQ} as the heavy-quark pole mass.
This, however, would not be a practical choice. The reason is that \eq{bare2pole:lattice}, 
which relates the pole mass with the bare lattice mass,  
does not converge in perturbation theory being plagued by the presence of an infrared renormalon of order~$\Lambda_{\rm QCD}$.
A solution consists in subtracting from the pole mass its infrared sensitive part and defining in this way a new, renormalon free, mass. 
Hence, the second step in order to use \eq{PS_Mass:HQET:mQ} as a fitting function for lattice data 
consists in interpreting the heavy-quark mass and the LECs in a renormalon-free subtraction scheme.
Several subtraction schemes have been suggested over the years. 
We adopt the so-called renormalon-subtracted (RS) scheme~\cite{Pineda:2001zq}, 
where the renormalon-subtracted mass is defined by subtracting the leading infrared renormalon from the pole mass. 
The relation of the RS mass with the $\MSbar$ mass is as accurate as the relation between the pole mass and the $\MSbar$ mass, i.e., 
it is known up to order $\alpha_s^4$. 
From \eq{MSbarratio2bareratio:lattice:hc} we may then relate the RS mass with the bare lattice mass.
In the expression of the RS mass, finite charm-mass effects have been taken into account up to two loops,  
which is consistent with the fact that we compare with data obtained from ensembles with $n_f=2+1+1$ flavors.
We use three active flavors in the analytical expressions~\cite{Ayala:2014yxa,Ayala:2016sdn}. 

Finally, the fitting function for the heavy-light meson masses that we use is  
\begin{eqnarray}
    M_H  &=&  m_h^\RS + \bar{\Lambda}^\RS + \frac{\mu_\pi^2 - \mu_G^2(m_h^\RS)}{2m_h^\RS}  
    + \frac{\rho}{(2m_h^\RS)^2}  \, , \label{eq:PS_Mass:HQET:mRS}
\end{eqnarray}
where $m_h^\RS$ is the RS mass of the heavy quark $h$, related to the bare lattice mass through \eq{MSbarratio2bareratio:lattice:hc} 
and the four-loop relation between the RS mass and the $\MSbar$ mass. The quantity $\rho$ parameterizes higher-order LECs.
The fit determines six free parameters: $\bar{\Lambda}^\RS$, $\mu_\pi^2$, $\mu_G^2(m_b^\RS)$, $\rho$, 
and through \eq{MSbarratio2bareratio:lattice:hc} $\mcstbar$ and $K_1$.
We set the prior value of $\mu_G^2(m_b^\RS)$ on the hyperfine splitting of $M_{B^*}-M_B$. 

\begin{figure}[bp]
\begin{tabular}{c c}
\includegraphics[width=0.48\textwidth]{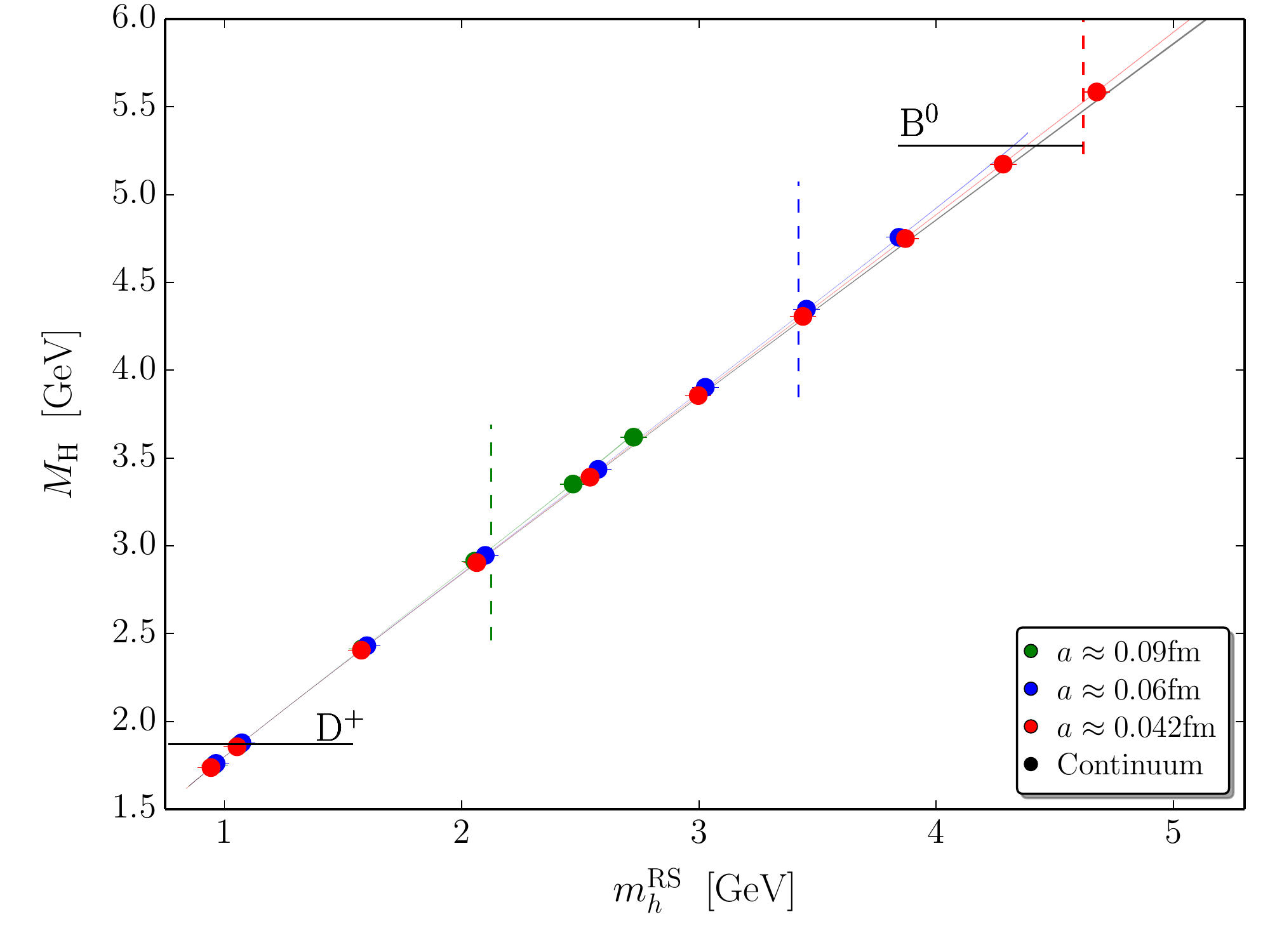} &
\includegraphics[width=0.48\textwidth]{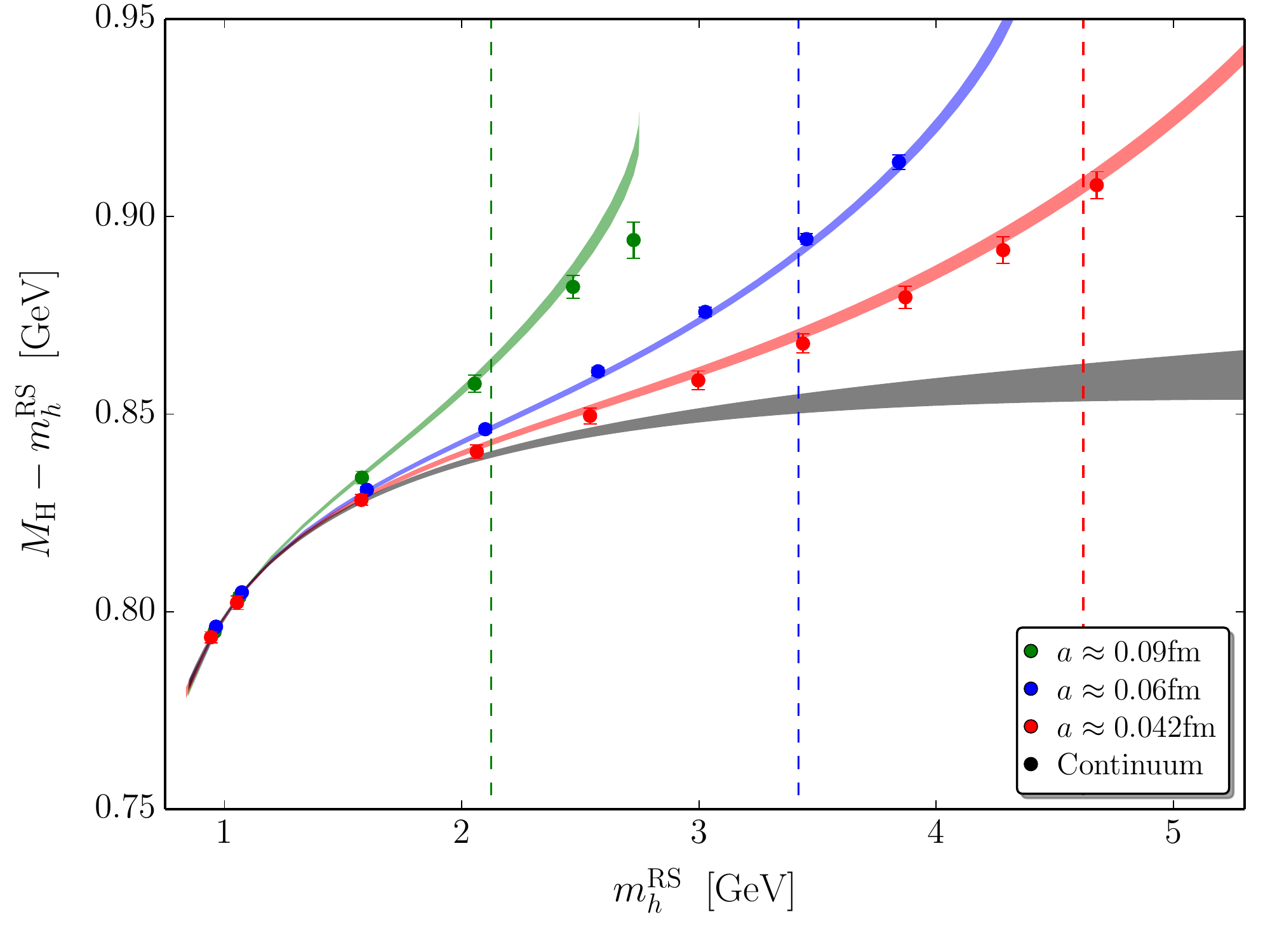}
\end{tabular}
\vspace{-5mm}
\caption{\label{fig:Masses} 
HQET fit of heavy-light-meson-mass data with valence heavy-quark masses $am'_c \le am_h <0.9$.
The dashed vertical lines indicate the cut $am_h = 0.9$ for each lattice spacing.
On the horizontal axis is the continuum limit of the $h$-quark RS mass. }
\vspace{-2mm}
\end{figure}

Figure~\ref{fig:Masses} illustrates a sample fit based on the method presented here. 
It shows qualitatively that the lattice artifacts are well modeled by our parameterization of the discretization errors. 
Further, discretization errors are small.
From the continuum extrapolation, we obtain the meson mass as a function of the heavy-quark mass in the RS scheme 
that may be eventually converted in the conventional $\MSbar$ scheme.
Hence, by looking at the $D$ and $B$ meson masses, we can determine the charm- and bottom-quark masses. 
Quantitative results will be presented in a future paper. 

\vspace{-2mm}
\section{Conclusion} 
\label{sec:Conclusion}
\vspace{-2mm}
We have presented the status of our analysis of $f_B$ and $f_{B_s}$ from a lattice-QCD calculation with 
the HISQ action for all quarks.
We anticipate that our calculations when completed will be the most precise to date. 
We also presented our method of extracting charm- and bottom-quark masses from heavy-light meson masses. 
This method also yields the the HQET quantities $\Lambdabar$ and $\mu_\pi^2$. 

\vspace{-3mm}
\acknowledgments
\vspace{-3mm}
Computations for this work were carried out with resources provided by the USQCD Collaboration; 
by the ALCF and NERSC, which are funded by the U.S. Department of Energy (DOE); 
and by NCAR, NCSA, NICS, TACC, and Blue Waters, which are funded through the U.S. National Science Foundation (NSF). 
Authors of this work were also supported in part through individual grants by the DOE and NSF (U.S); 
by MICINN and the Junta de Andaluc\'ia (Spain); by the European Commission; and by the German Excellence Initiative.


\end{document}